\documentclass[%
 aip,
jcp,%
 amsmath,amssymb,
 reprint,%
]{revtex4-1}

\usepackage{dcolumn}
\usepackage{bm}
\usepackage{graphicx}
\usepackage[space]{grffile}
\usepackage{bbm}
\graphicspath{{"../orthogonal dyad_magic_angle/figs/"}}

\begin{document}

\title[]{Geometric dependencies of vibronically mediated excitation transfer in rylene dyads }

\author{V\'{a}clav Perl\'{\i}k }
\thanks{These authors contributed equally}
\author{Vladislav Sl\'{a}ma }
\thanks{These authors contributed equally}
\affiliation{Institute of Physics, Faculty of Mathematics and Physics, Charles University, Ke Karlovu 5, Prague, 121 16 Czech Republic}
\author{Craig N. Lincoln}
\affiliation{Photonics Institute, TU Wien,  Gu{\ss}hausstra{\ss}e 27-29, 1040 Vienna, Austria }
\author{Heinz Langhals}
\affiliation{Department of Chemistry,
 Ludwig-Maximilians-Universit\"{a}t M\"{u}nchen,
Butenandtstra{\ss}e  5-13,  D-81377 Munich, Germany}
\author{Eberhard Riedle}
\affiliation{Lehrstuhl f{\"u}r BioMolekulare Optik, Ludwig-Maximilians-Universit\"{a}t M\"{u}nchen, Oettingenstra{\ss}e 67, 80538 Munich, Germany}
\author{Tom\'{a}\v{s} Man\v{c}al}
\affiliation{Institute of Physics, Faculty of Mathematics and Physics, Charles University, Ke Karlovu 5, Prague, 121 16 Czech Republic}
\author{J{\"u}rgen Hauer}
\affiliation{Photonics Institute, TU Wien,  Gu{\ss}hausstra{\ss}e 27-29, 1040 Vienna, Austria }
\affiliation{Professur f\"{u}r Dynamische Spektroskopien, Fakult\"{a}t f\"{u}r Chemie, Technische Universit\"{a}t M\"{u}nchen, Lichtenbergstr. 4, D- 85748, Garching b. M\"{u}nchen, Germany}
\author{Franti\v{s}ek \v{S}anda}%
\email{sanda@karlov.mff.cuni.cz}
\affiliation{Institute of Physics, Faculty of Mathematics and Physics, Charles University, Ke Karlovu 5, Prague, 121 16 Czech Republic}

\date{\today}

\begin{abstract}
We study the excitation transfer  in various geometric arrangements of rylene dimers using absorption, fluorescence and transient absorption spectra. Polarization and detection frequency dependencies of transient absorption track the interplay of transfer and vibrational relaxation within the dyads.
We have resolved microscopic parametrization of intermolecular coupling between rylenes and reproduced  transport data.
Dynamical sampling of molecular geometries captures thermal fluctuations for Quantum Chemical estimate of couplings for orthogonally arranged dyad,
where static estimates vanish and normal mode analysis of fluctuations underestimates them by an order of magnitude.
Nonperturbative accounts for the modulation of transport by strongly coupled anharmonic vibrational modes
is provided by a vibronic dimer model. Vibronic dynamics is demonstrated to cover both the
F\"{o}rster transport regime of orthogonally arranged dyads and the strong coupling regime of parallel chromophores and allows us to model signal variations along the detection frequency.
\end{abstract}
\maketitle

\section{Introduction}
\label{s:Introduction}

Rylene dyes have attracted attention as a suitable toy model for studying excitation energy transfer (EET) due to their versatility and convenient spectroscopic properties. These molecules show high fluorescence quantum yield and photostability \cite{langhals1995cyclic} and were thus applied as laser dyes\cite{lohmannsroben1989laser,qian2003photostability}, fluorescent light collectors\cite{seybold1989new}, fluorescent probes\cite{bo2013assembly} or fluorophores for single-molecule spectroscopy\cite{mais1997terrylenediimide,lang2005photophysical}. Their significant charge transport abilities \cite{huang2011perylene} can be extended over larger aggregates, so they are also proposed as a building blocks for organic photovoltaics\cite{hofmann2010mutual,holcombe2011steric}.
The family of rylenes, e.g.,  perylene, terylene and their chemical derivatives,
 shows  similar optical properties\cite{herrmann2006industrial}, fluorescence lifetimes around 5ns with quantum yield  near unity
and  absorption/fluorescence spectra dominated by
a central transition between HOMO and LUMO state in the visible spectral region modulated by a strong ring stretching vibrational mode around 1300-1400 cm$^{-1}$ \cite{ambrosino1965vibrational}.
Rylenes can be linked into dyads whose spectral features and  excitation energy transfer is fine-tuned by substituents in the side position \cite{fron2008photophysical,osswald2007effects} and geometric arrangements. These variations shall be understood as different parametrizations of an abstract model of rylene dyad.

Still, the broad variability of geometric arrangements of rylenes in  dyads implies large variations in magnitude of intermolecular coupling. Accordingly, transport interpolates between regimes \cite{wurthner2001fluorescent} of weak intermolecular coupling described by F\"{o}rster transfer theory\cite{forster1948zwischenmolekulare}  to regimes where the intermolecular coupling  dominates over electron-vibrational modulations (Redfield theory\cite{redfield1957theory}). General efforts to unify the two theoretical frameworks include parametric interpolations between these limits by  polaron transforms \cite{zimanyi2012theoretical,fujihashi2013improved} or explicit treatments of vibrations.
For the latter instance, strong spectral diffusion of excitons can be modelled by modulating excitonic energies by a stochastic Markovian coordinate representing effect of overdamped vibrations \cite{sanda2006cooperative,sanda2008stochastic,tanimura2006stochastic}. The pronounced vibronic progressions observed in rylenes imply yet another approach to vibrations; to treat a dominant high frequency vibrational modes on equal footing with electronic degrees of freedom introducing thus structures of vibronic levels. The effects of the remaining vibrations and the solvent are already moderate and their perturbative treatment can be justified for a broad family of rylene dyadic systems.
The outlined vibronic approach is not limited to predicting the rates of energy transfer. It describes also the full optical dynamics and accounts
for variations of the spectral dynamics along  excitation and detection frequencies
and tracks excitation transfer through the energy ladder.

\begin{figure}[]
  \begin{center}
    \includegraphics[width=\linewidth]{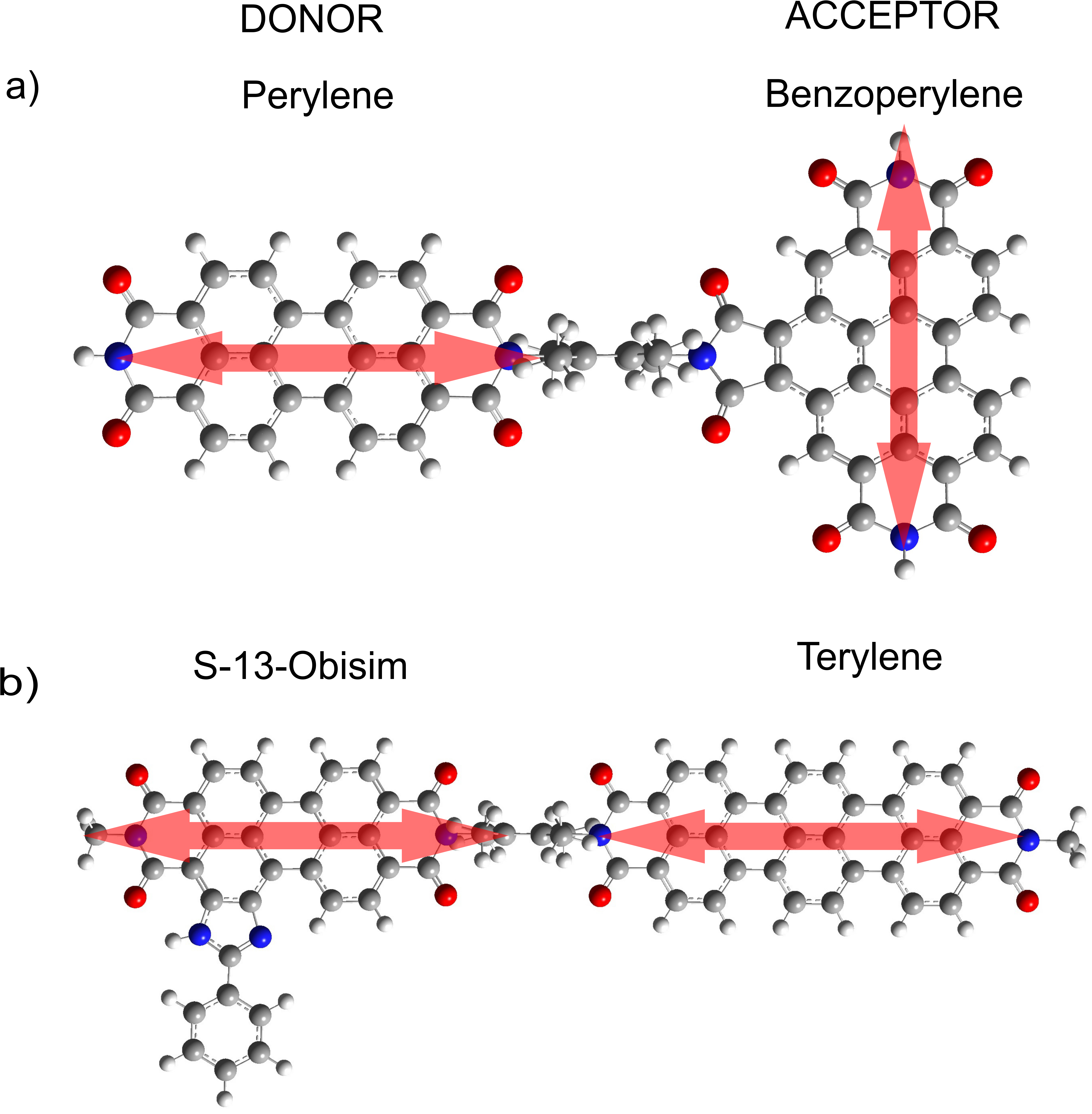}
  \caption{
  Donor-acceptor systems used in experiments.  a) Perylene-benzoperylene bisimide dyad in orthogonal arrangement. b) S13-obisim-terylene dyad in parallel arrangement dipole moments between the donor (S-13-obisim) and the acceptor (terylene).}
    \label{f:molecules}
  \end{center}
\end{figure}

In the present work, femtosecond transient absorption measurements are used to study
excitation transport  in rylene dyads of two distinct geometries.
In the first arrangement,  the transition dipoles of the donor (perylene) and acceptor (benzoperylene) are strictly orthogonal at zero temperature    (Fig. \ref{f:molecules}a) with (prima facie) expectations of zero interchromophoric coupling and no transport.  Fast transport reported in Ref \onlinecite{langhals2010forster} has thus been hypothetically attributed to thermal fluctuations from the  orthogonal geometry, supported by experimental evidence of the transfer dependencies on temperature or solvent \cite{langhals2010forster,nalbach2012noise}, however, the quantitative microscopic treatment was unsuccessful so far.
Fully {\it ab initio} parameterizations are far too ambitious, but even when quantum chemical calculation is limited to the interchromophoric couplings
(and the electron-vibrational and solvent couplings fitted on the absorption spectra of constituents), the transport is underestimated by an order of magnitude.
Here, we aim to prove the F\"{o}rster picture of transport  by checking  delay time and detection frequency and polarization dependencies of the transient absorption signal, and reproducing it with a vibronic dynamical model.
To this end, we have to revisit microscopic density functional theory (DFT) parametrization of the coupling by sampling the molecular geometries using molecular dynamics (MD) and by a careful statistical analysis of the predicted coupling distributions.

In the second studied geometric arrangement, the transition dipoles of the terylene and  S-13-obisim constituents are  parallel (Fig. \ref{f:molecules}b), which yields strong coupling far beyond the applicability of F\"{o}rster transport theory. This will be demonstrated by  no-additivity of absorption spectra and detection dependencies of transient absorption dynamics.
Theoretical model should decipher the complex dynamics and detection frequency dependencies in transient absorption data.

\section{Methods}
\label{s:Methods}

\subsection{Estimates  of intermolecular  coupling}
\label{s:Estimates}
Optical dynamics of dyads is induced by interaction ($\hat{V}_{AD}$) between the electronic structures of donor ($\hat{H}_{el}^{D}$) and acceptor ($\hat{H}_{el}^{A}$)
and is further modulated  by dynamics of nuclei ($\hat{H}_{nc}$)
\begin{equation}\label{}
\hat{H} = \hat{H}_{el}^{A} + \hat{H}_{el}^{D} + \hat{H}_{nc} + \hat{V}_{AD}.
\end{equation}
In the Born-Oppenheimer  approximation electronic and nuclear motions are separated.  In the following we refer to the local instantaneous electronic states ($k=0,1,,\dots$)
\begin{equation}\label{eq:time_independent_schre}
\hat{H}_{el, \bm{R}}^A \Psi_{k,\bm{R}}^A(\bm{r}_A) = E_{k,\bm{R}}^A, \Psi_{k,\bm{R}}^A(\bm{r}_A)
\end{equation}
where $\bm{r}_A = (r_1,r_2,\ldots)$ is  a collection of electronic coordinates on the acceptor,
and the parameter $\bm{R}$ stands for molecular geometry specified by collection of nuclear coordinates.
The complete information of many-body wavefunction is excessively abundant, and for most purposes shall be reduced to the one-particle transition densities between the  electronic states $k$ and $k'$ defined as
\begin{equation}\label{eq:transition_density}
\begin{split}
& \rho^A_{kk',\bm{R}}(r) = \int \Psi^A_{k,\bm{R}}(r_1,r_2,\ldots) \Psi_{k',\bm{R}}^{*A}(r_1,r_2,\ldots)  \\
& \qquad \qquad \times \sum_{j \in \{A\}} \delta(r-r_j)  \prod_{i \in \{A\}} dr_i .
\end{split}\end{equation}
The donor variables $E^{D}$, $\bm{r}_D$, $\Psi^D$, $\rho^A$ are defined analogically.

Interchomophoric coupling accounts for electrostatic interaction of electronic $r_i$  coordinates of acceptor ($i \in \{A\}$) with that of donor ($j \in \{D\}$), for their interaction with nuclear coordinates $R_J$ (of proton number $Z_J$) outside the acceptor (can include linker), and vice versa
\begin{equation}\label{}
\hat{V}_{AD} = \sum_{\substack{i \in \{A\} \\ j \in \{D\} }}\frac{1}{|r_i-r_j|} - \sum_{\substack{i \in \{A\} \\ J \notin \{A\}}}\frac{Z_J}{|r_i-R_J|}
- \sum_{\substack{I \notin \{D\} \\ j \in \{D\} }}\frac{Z_I}{|R_I-r_j|}.
\end{equation}
Intermolecular exciton transfer occurs predominantly between the two lowest excited states $|\Psi_{1,\bm{R}}^A \rangle |\Psi_{0,\bm{R}}^D \rangle$ and $|\Psi_{0,\bm{R}}^A \rangle |\Psi_{1,\bm{R}}^D \rangle$.
Resonant coupling between them
\begin{equation}\label{eq:resonant_interaction}
J_{\bm{R}} \equiv \langle \Psi^A_{1,\bm{R}}| \langle \Psi^D_{0,\bm{R}}| \hat{V}_{AD}|\Psi^A_{0,\bm{R}}\rangle |\Psi^D_{1,\bm{R}}\rangle
\end{equation}
 can be recast in terms of transition densities as
\begin{equation}\label{eq:tdc}
J_{\bm{R}} = \int \frac{\rho^A_{01,\bm{R}}(r_a)\rho^D_{01,\bm{R}}(r_d)}{|r_a - r_d|} dr_a dr_d .
\end{equation}
The direct discretization of Eq. (\ref{eq:tdc})  known as the transition density cube (TDC) method \cite{krueger1998calculation} shall be used at short intermolecular distances.
When the molecules are far apart (with respect to molecular size), Eq. (\ref{eq:tdc}) is well approximated by classical dipole-dipole interaction where dipoles are $\vec{\mu}_{\bm{R}}^A \equiv \int \vec{r} \rho^A_{01,\bm{R}}(r) dr$.

At finite temperatures $1/k_B\beta$, electronic structures and couplings depend on
molecular geometry, which should be sampled
along Boltzmann distribution, predicting expected value $\langle f \rangle $ of the quantity $f$ to be
\begin{equation}
 \langle f \rangle \equiv  \frac{ \int f({\bm{R}})e^{-\beta E(\bm{R})}  {\rm d}\bm{R}}{ \int  e^{-\beta E(\bm{R})} {\rm d}\bm{R}}.
\label{eq:normal_mode_coupling}
\end{equation}
Here, we evaluate the ground state energies
\begin{equation}
E(\bm{R}) \equiv \langle \Psi^A_{0,\bm{R}}|\langle \Psi^B_{0,\bm{R}}|\hat{H}|\Psi^A_{0,\bm{R}}\rangle |\Psi^N_{0,\bm{R}}\rangle
\label{eq:ground_state_energy}
\end{equation}
using density functional theory (DFT) which draws upon recasting Eq. (\ref{eq:ground_state_energy})
in the form of Kohn-Sham functional \cite{parr1989density}.
In particular, the ground state energies (Eq. (\ref{eq:ground_state_energy})) were  calculated with B3LYP DFT functional in 6-311G(p,d) basis. Similarly the transition densities $\rho_{01}$ were obtained from TD-DFT using long range corrected CAM-B3LYP functional.

The coupling estimates are usually feasible at zero temperature geometry $\bm{R}_0$, characterized by the minimal  ground state energy $E(\bm{R}_0)\le E(\bm{R})$.
For orthogonal dyad, however,  $J_{\bm{R}_0} \approx 0$,  and the thermal fluctuations might dominate the transport. We thus explore static distributions
of coupling $\langle \delta (J- J_{\bm{R}})\rangle$. To avoid costly complete exploration of high dimensional ${\bm{R}}$ configuration space, we compare sampling by normal mode analysis (NMA) and by molecular dynamics (MD).

In the NMA, energy  is expanded to second order $E({\bm R} ) = E({\bm R}_0) + \sum_{I,J}{\mathcal E }_{IJ} (R_I-R_{0,I})( R_J-R_{0,J})$ around the minimum. The Hessian ${\mathcal E}_{IJ} =  \frac{\partial E}{\partial R_I \partial R_J} $ is diagonalized to obtain the normal mode coordinates $Q_I$.
Geometry sampling is restricted along a single normal coordinate $Q_I$, otherwise weighted according to
Eq. (\ref{eq:normal_mode_coupling}), and the couplings on the samples are evaluated using TDC (Eq. (\ref{eq:tdc})).

However, the NMA ignores the complexity of potential surfaces, and also usually neglects solvent effects.
An alternative approach is thus to account for the solvent by employing  MD to sample molecular geometries, avoiding also the harmonic approximation to $E(\bm{R})$ inherent to NMA. Using AMBER  package with GAFF force field and RESP charges calculated by Gaussian09 software we equilibrated the molecule in toluene at a room temperature (300K) for 10ns,
and then sampled $N=1000$ molecular geometries along 40 ps MD trajectory. 

\subsection{Vibronic model for rylene dyads}
\label{s:Vibronic_model}
Now we will connect the microscopic parameterizations of the previous section to the vibronic dynamics of Ref \onlinecite{perlik2017vibrational}. To this end, we first define excitonic states, formally fixing the wavefunction at some typical geometry, say  $\bm{R}_0$.
For the transient absorption spectroscopy, the ground state $|g\rangle \equiv |\Psi^A_{0,\bm{R}_0}\rangle|\Psi^D_{0,\bm{R}_0}\rangle$,  the two singly excited states $|e_A\rangle \equiv  |\Psi^A_{1,\bm{R}_0}\rangle|\Psi^D_{0,\bm{R}_0}\rangle$ and $|e_D\rangle \equiv  |\Psi^A_{0,\bm{R}_0}\rangle|\Psi^D_{1,\bm{R}_0}\rangle$,  and the three doubly excited states
$|f_A\rangle \equiv |\Psi^A_{2,\bm{R}_0}\rangle |\Psi^D_{0,\bm{R}_0}\rangle $ ,
$|f_D\rangle \equiv |\Psi^A_{0,\bm{R}_0}\rangle |\Psi^D_{2,\bm{R}_0}\rangle$, and
$|f\rangle \equiv |\Psi^A_{1,\bm{R}_0}\rangle |\Psi^D_{1,\bm{R}_0}\rangle$
are relevant.  Similarly, we fix transition frequencies at typical values, formally setting $\epsilon^{A(D)} \equiv E^{A(D)}_{1,\bm{R}_0}-E^{A(D)}_{0,\bm{R}_0}$ for acceptor (donor) excitation frequency, $\epsilon_f^{A(D)} \equiv E^{A(D)}_{2,\bm{R}_0}-E^{A(D)}_{0,\bm{R}_0}$ for double excitation.  The Frenkel exciton Hamiltonian thus reads
\begin{equation}
\begin{split}
\hat{H}_{el} &= \epsilon^{A} |e_A \rangle\langle e_A| + \epsilon^{D} |e_D \rangle\langle e_D|+ J(|e_A\rangle\langle e_D| + |e_D\rangle \langle e_A|)\\
&+ \epsilon^{A}_f |f_A \rangle\langle f_A| + \epsilon^{D}_f  |f_D \rangle\langle f_D|+ (\epsilon^{A} +\epsilon^{D}) |f \rangle\langle f|,
\end{split}
\label{eq:frenkel_hamiltonian}
\end{equation}
where $J$ is effective value of the intermolecular coupling between $e_A$ and $e_D$ (other elements of $\hat{V}_{AD}$ were neglected).
In most situations, thermal fluctuations are minor $|\langle J^2\rangle - \langle J\rangle^2|\ll \langle J\rangle ^2$ and the effective coupling will be represented by mean value often corresponding to that of ${\bf R}_0$ geometry $J\equiv \langle J\rangle \approx  J_{{\bf R}_0}$.  In the opposite case of dominating fluctuations $\langle J\rangle <  \sqrt{\langle J^2\rangle}$ (applicable to orthogonal dyad) the coupling will be represented
in the excitonic model by the second moment of distribution $J\equiv \sqrt{\langle J^2\rangle}$, as most transport theories scale rates as $\propto J^2$.

We next introduce vibronic dynamics, which allows us a unified treatment of the transport in both types of dyads.
It arises when  underdamped vibrations modulating the Frenkel excitons ($\hat{H}_{vib}^A$, $\hat{H}_{vib}^D$)
are separated from $\hat{H}_{nc}$ and included into the system Hamiltonian $\hat{H}_s$
\begin{equation}
\hat{H}_s = \hat{H}_{el} + \hat{H}_{vib}^{A}\otimes\mathbbm{1}^D + \mathbbm{1}^A\otimes\hat{H}_{vib}^{D}.
\label{eq:vibronic_hamiltonian}
\end{equation}
In particular, we adopt the anharmonic oscillator model  $V(q)=\frac{1}{2} m \omega^2 \hat{q}^2 +\alpha\hat{q}^3$ of Ref. \onlinecite{galestianpour2017anharmonic}  to represent typical ring stretching mode around 1400 cm$^{-1}$. For instance, for the acceptor we have
\begin{equation}
\hat{H}_{vib}^{A} =  \frac{\hat{p}_A^2}{2m} + \sum_{{\rm j}=0}^{2} V(q_A-d^{A}_{\rm j}) |\Psi^A_{{\rm j},\bm{R}_0} \rangle\langle \Psi^A_{{\rm j},\bm{R}_0}|,
\label{eq:vibrational_hamiltonian}
\end{equation}
where $d^{A}_{\rm j}$ is displacement of j-th electronic surface.

The dynamical modulation by other  nuclear coordinates(including solvent) is represented  by
the quantity $\lambda_V$ ($\lambda_W$) and the relaxation rates $\Lambda_V$ ($\Lambda_W$) for linear (quadratic) vibration-to-bath coupling \cite{perlik2017vibrational},
and $\lambda_A$ ($\Lambda_A$)   for electronic dephasings. 
Parametrizing them by expanding the eigenspectrum Eq. (\ref{eq:time_independent_schre}) or quantum chemistry calculations
is possible but challenging task \cite{olbrich2010time,olbrich2011theory}, we follow the common practice of fitting these parameters from absorption spectrum.

The molecular system is probed by laser fields as described by interaction Hamiltonian
\begin{equation}
\label{eq:interaction_hamiltonian}
\begin{split}
\hat{H}_{i} &= \mu_A(t)|e_A\rangle\langle g|  +\mu_D(t)|e_D\rangle\langle g|  + \mu_{fA}|f\rangle\langle e_A| \\
&+ \mu_{fD}|f\rangle\langle e_D| + \nu_A(t)|f_A\rangle\langle e_A| + \nu_D(t)|f_D\rangle\langle e_D|  + {\rm h.c.},
\end{split}
\end{equation}
where $\mu_{A(D)}(t) \equiv  \vec{\mu}_{A(D)} \cdot \vec{E}(t)$ and $\nu_{A(D)}(t) \equiv \vec{\nu}_{A(D)} \cdot \vec{E}(t)$ are the projections of laser field $\vec{E}(t)$ on the transition dipoles $\vec{\mu}_{A}(\vec{\mu}_{D})$ and $\vec{\nu}_{A}(\vec{\nu}_{D}$) between the $g$  and $e$ states and between the $e$  and $f$  states of the acceptor (donor) at ${\bf R}_0$ geometry, respectively.

The linear and nonlinear optical response is simulated using machinery of Ref \onlinecite{perlik2017vibrational},
i.e. by combining quantum master equation to describe vibronic population transfer and second cumulants for line-shapes.
 We also adopted  factors $\Omega$,  and $\Omega^3$
 to connect (transient) absorption and fluorescence spectra, respectively, with the response functions along the Ref. \onlinecite{angulo2006recalling}.
Polarization dependencies of the response includes orientational averaging procedure described in Ref. \onlinecite{andrews1977three}.
We have also accounted for narrow band pump excitation\cite{perlik2017finite} using the actual pulse profiles.

\subsection{Sample preparation}
The preparation of the orthogonal dyad has been published previously\cite{langhals2008fret}.
The parallel was prepared as follows:
Terylene anhydride carboximide was solubilised by means of the long-chain secondary alkyl substituent 1-nonyldecyl at the nitrogen atom and condensed with an excess of
2,3,5,6-tetramethylbenzene-1,4-diamine. The free primary amino group of the thus obtained terrylenebiscarboximide was further condensed\ with an imidazoloperyleneanhydridecarboximde to obtain the dyad with a rigid, orthogonal spacer between the two chromophores for electronic decoupling, where the two peripheric sec-alkyl substituents render both a sufficiently high solubility and a low tendency for aggregation.

\subsection{Optical methods}
Broadband pump-probe measurements with a temporal resolution of about 50 fs were carried out in a setup described in detail previously\cite{megerle2009sub,riedle2013electronic}. Probing white light was generated by focusing the output from the Ti:Sapphire laser into a moving $\text{CaF}_{\text{2}}$ plate. The parallel dyad was excited with noncollinear optical parametric amplifier (NOPA) pulses centred at 550 nm and spectrally limited as to excite only the highest energy band (see Fig. 8). The orthogonal dyad was selectively excited at 435 nm with the frequency doubled NOPA output with about 30 fs pulse duration. At this wavelength, the perylene moiety (i.e. the energy acceptor) has negligible absorbance (see Fig. 3).

\section{Results}
\label{s:Results}

\subsection{Distributions of the intermolecular coupling}
\label{s:Coupling}

\begin{table}
\begin{tabular}{|c|c|c|c|}
  \hline
  Normal mode & Frequency [cm$^{-1}$] & $\sqrt{\langle J^2\rangle}$ \\
  1 & 7.93 &  1.55\\
  2 & 8.4 & 0.22\\
  3 & 12.39 & 0.09\\
  4 & 20.81 & 1.34\\
  5 & 21.98 & 0.02\\
  \hline
\end{tabular}
\caption{Five lowest normal modes and variances of coupling distributions $\langle J^2 \rangle $. In all cases $\langle J \rangle \approx 0$ .}
\label{t:normal_modes}
\end{table}

The DFT-optimized geometry $\bm{R}_0$ of the dyad displayed at Fig 1a has orthogonal dipoles and vanishing coupling $J_{\bm{R}_0} = 0$ within the numerical error. For NMA analysis we identified five lowest normal modes and calculated distributions of coupling at room temperature along each of these coordinates. Coupling distributions are approximately Gaussian with zero mean $\langle J\rangle \approx 0$. Variances $\langle J^2
 \rangle$ summarized in the table \ref{t:normal_modes} are  by an order of magnitude insufficient to recover the dynamics  of F\"{o}rster transport ($J\approx 17$cm$^{-1}$) reported in Ref. \onlinecite{langhals2010forster}.  No prominent vibration thus can be associated with coupling fluctuations, instead we noticed complex $E(\bm{R})$ landscapes outside the main axes, which may be responsible for larger couplings.

\begin{figure}[]
  \begin{center}
  \includegraphics[width=0.49\linewidth]{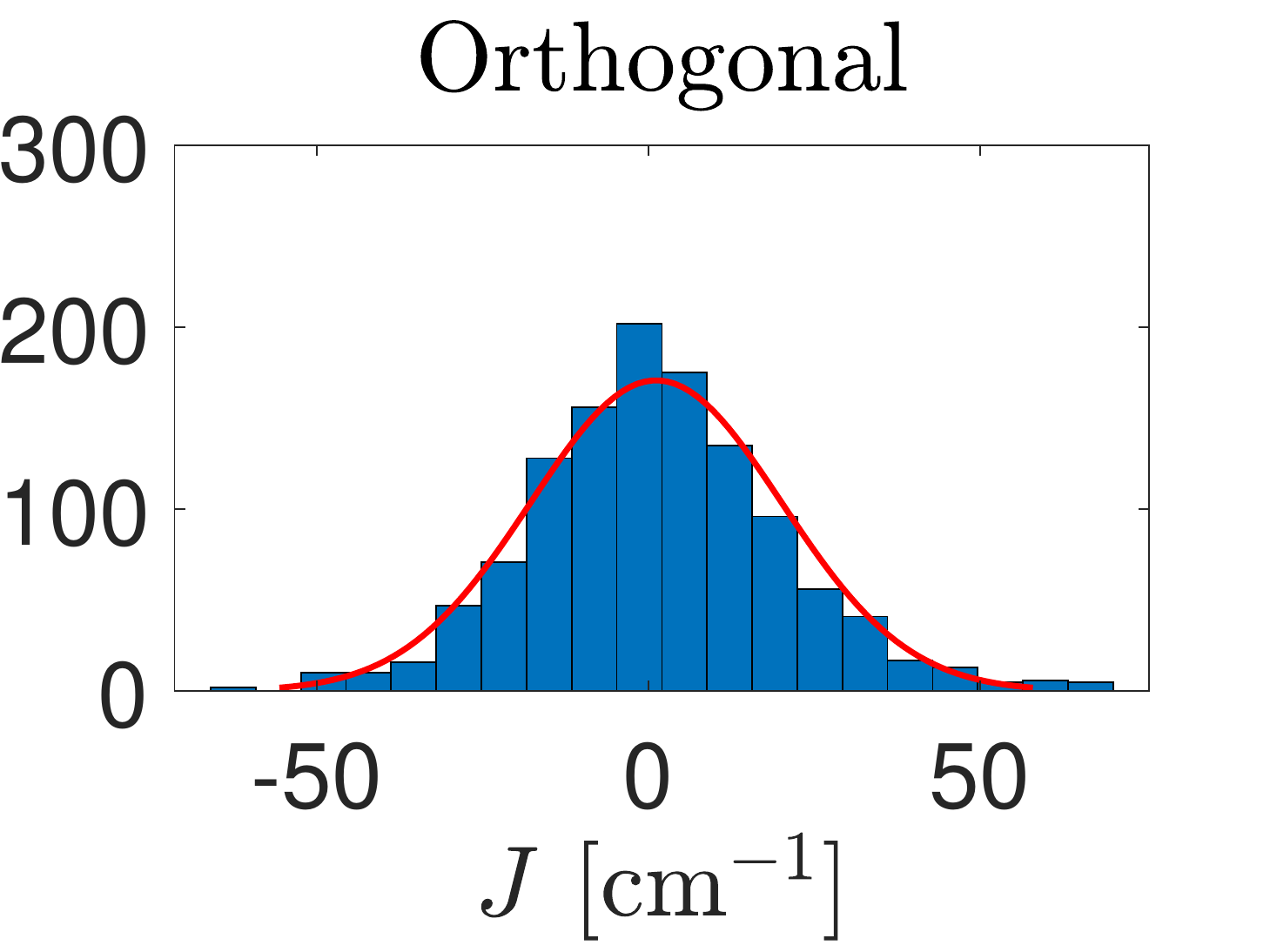}
   \includegraphics[width=0.49\linewidth]{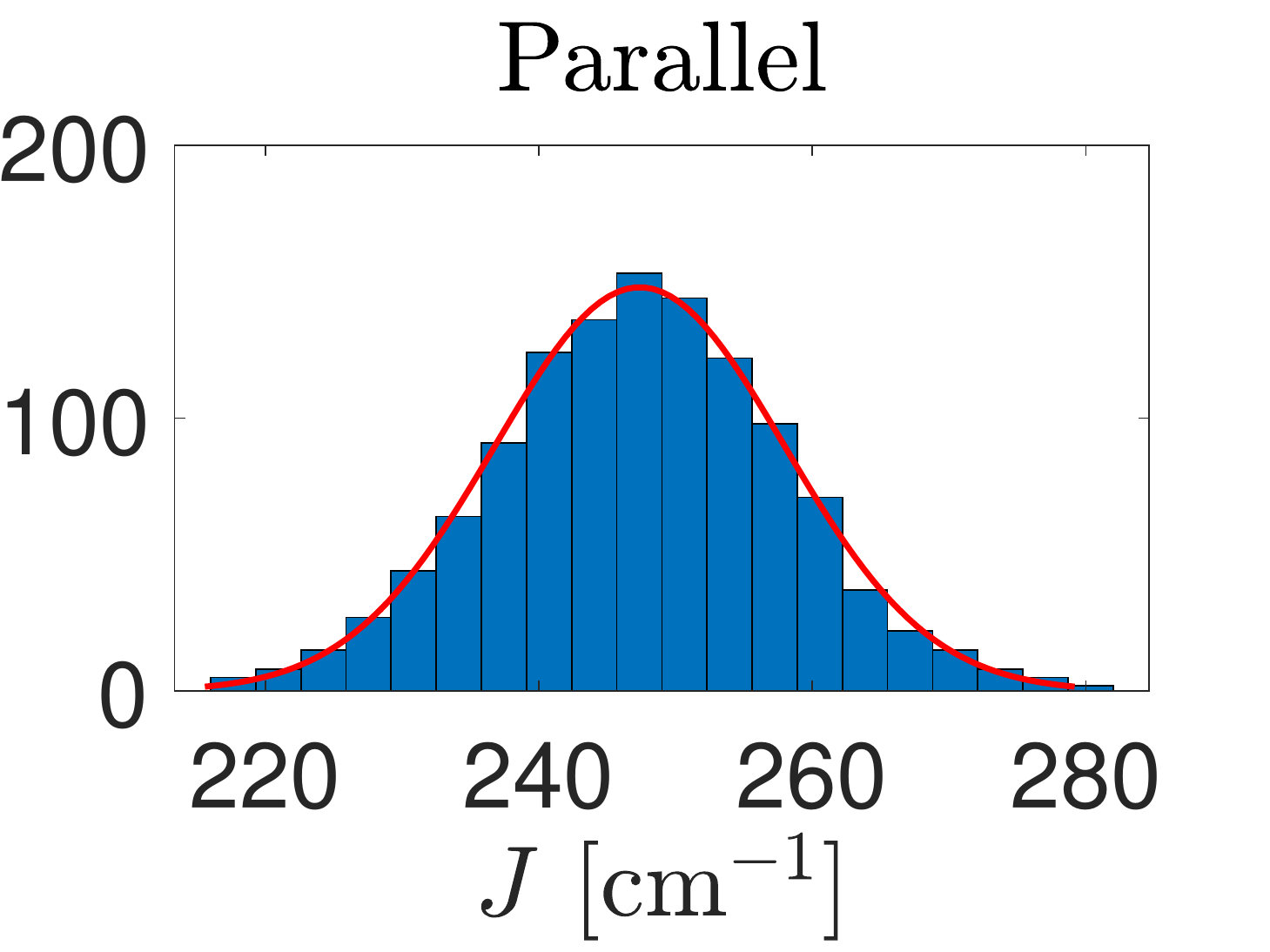}
  \caption{Histogram of coupling : left panel, Orthogonal dyad dissolved in toluene,  mean and variance are $\langle J \rangle = 0.8$, $\sqrt{\langle J ^2 \rangle - \langle J \rangle^2} = 19$. Right panel: parallel dyad and $\langle J \rangle = 247$, $\sqrt{\langle J ^2 \rangle - \langle J \rangle^2} = 11$.}
    \label{f:histogram}
  \end{center}
\end{figure}

Regular samplings of (high dimensional) configuration space are ineffective. We have thus used MD to sample geometries,
 also allowing us to account for toluene as a solvent.
 Using AMBER  software package,  geometries were sampled with 2fs timestep  after initial  10 ns equilibration at room temperature (300K). For each geometry, coupling was estimated using the methods of Section \ref{s:Estimates}. Distributions thus obtained are symmetric around $J=0$ within statistical error  with variance $\sqrt{\langle J^2\rangle} = 19\;{\rm cm}^{-1}$ .
We also noticed that distribution may deviate from standard Gaussian statistics, and performed  deeper statistical analysis of the fluctuating coupling. Lilliefors test \cite{lilliefors1967kolmogorov} found significant  difference from normal (Gaussian) distribution (at confidence level 0.05),  which we ascribe to the complex landscapes diverting far from harmonic vibrational motions.

We have repeated the calculations also for the parallel dyad (Fig. \ref{f:molecules} b) keeping the same technical settings. The dipoles of donor and acceptor are parallel in optimal geometry $\bm{R}_0$. The coupling of $J= 235$ cm$^{-1}$ obtained using TCD method (Eq. (\ref{eq:tdc})) is somewhat stronger than the one predicted in the dipole approximation ($J= 150$cm$^{-1}$; both at $\bm{R}_0$).
The distributions of coupling (MD sample of N=1000 geometries)  at Fig \ref{f:histogram}b are narrowly peaked at $J=247$cm$^{-1}$, (variance  $\sqrt{\langle J^2\rangle -\langle J\rangle^2}=11$cm$^{-1}$)
and without  significant differences from Gaussian distribution (Lilliefors test).
The contribution of fluctuations is thus insignificant for parallel dyad, i.e. the distribution has a rather small variance with respect to the mean, and the prediction of mean is not different from the coupling obtained by TCD for optimal geometry $\bm{R}_0$.

\subsection{Spectra of absorption and fluorescence}
\label{s:Experiment}
\begin{table}
\begin{tabular}{|c|c|c|c|c|}
  \hline
 molecule & perylene & benzoperylene & S-13-obisim  & terylene \\ \hline
  dyad & orthogonal & orthogonal & parallel & parallel \\ \hline
  role  & donor & acceptor &donor & acceptor   \\  \hline
  $\omega$ &  1470 & 1400&1383 & 1377 \\  \hline
   $d^2m\omega/(2\hbar)$ & 0.8& 0.7& 0.6& 0.48 \\  \hline
  $\alpha$ & -35 & -10 & -14 & 7 \\  \hline
  $E_1-E_0$ & 21650+ & 19100 & 17000& 15400 \\  \hline
$\Lambda_V$ & 200 & 266 & 60 & 5 \\  \hline
$\Lambda_A$ & 200 & 257 & 150& 150\\  \hline
$\Lambda_W$ & -  & - & 15 & 15 \\ \hline
$\lambda_V$ & 200  & 270 & 300& 140 \\  \hline
$\lambda_A$ & 350 & 265 & 250& 300\\  \hline
$\lambda_W$ & 0 & 0 & 15 & 15\\  \hline
$ \mu$ &  0.83 & 1 & 0.83 & 0.9 \\
  \hline
\end{tabular}
\caption{Parametrization (in cm$^{-1}$) of the vibronic model for the absorption and fluorescence spectra of the dyads constituents.}
\label{t:parametrization}
\end{table}
\begin{figure}[]
  \begin{center}
    \includegraphics[width=0.49\linewidth]{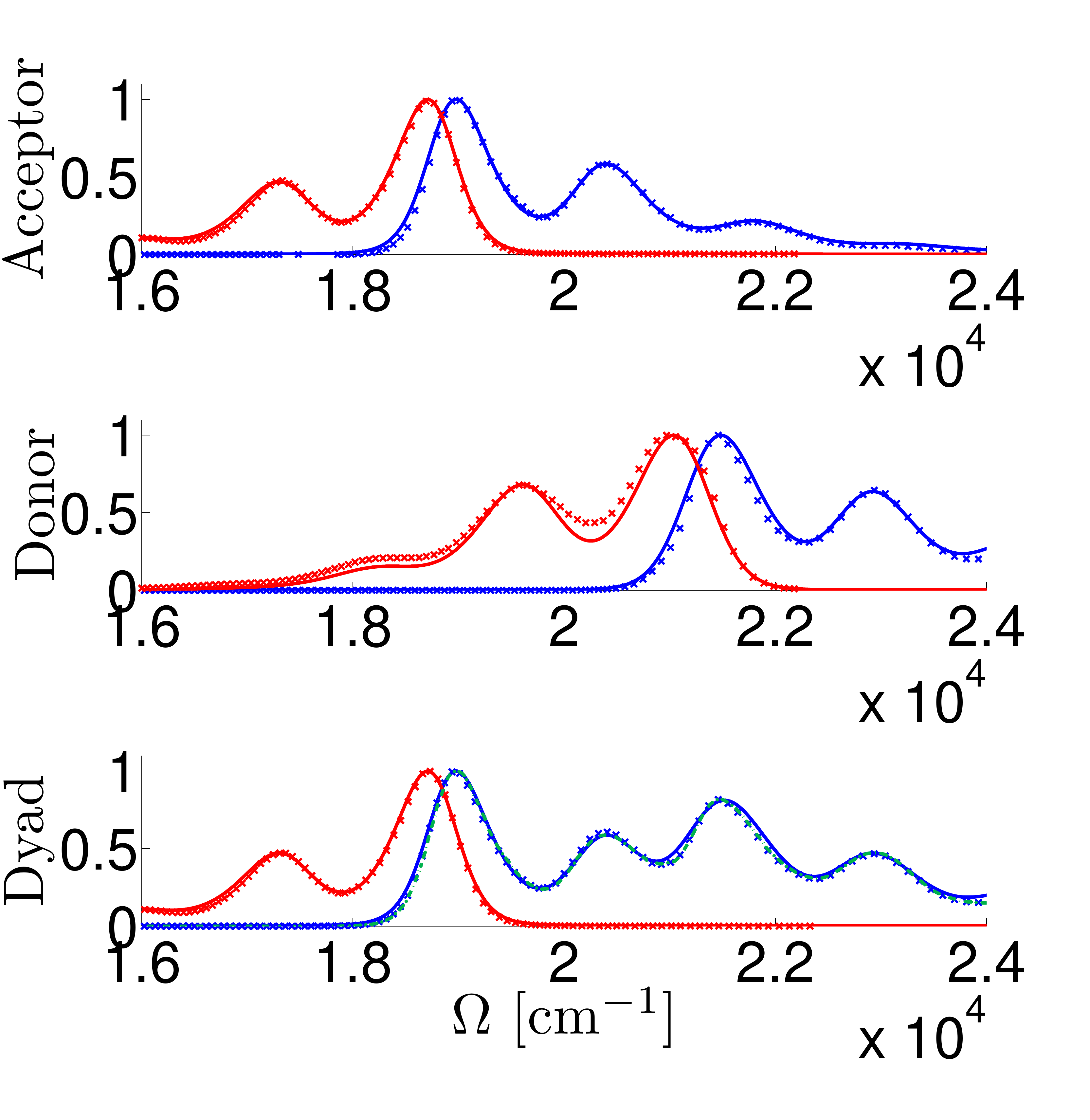}
     \includegraphics[width=0.49\linewidth]{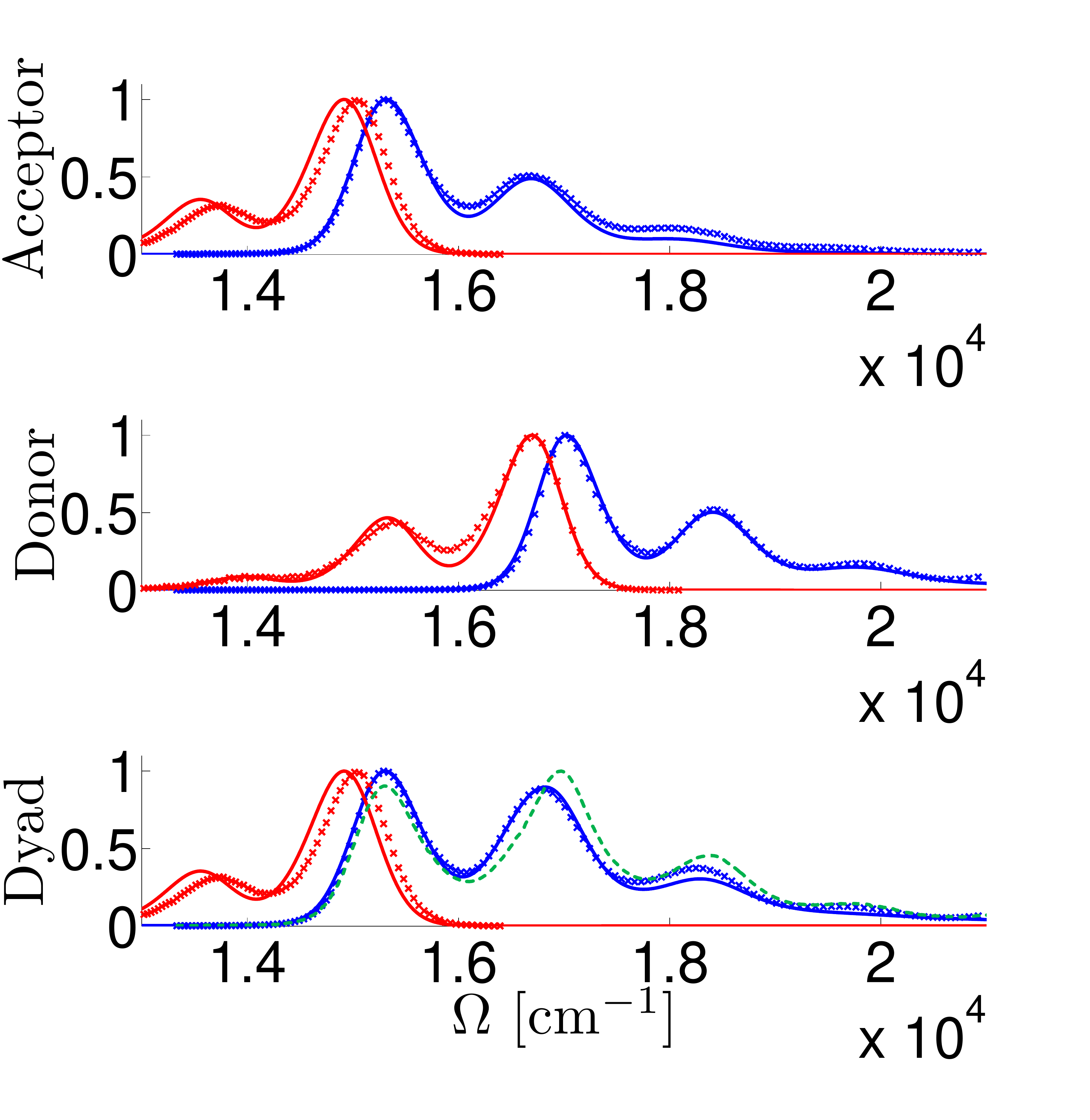}
      \caption{Experimental (crosses) and simulated (solid
lines) absorption (red lines) and fluorescence (blue lines) spectra for acceptor (top panels), donor (central panel), and dyad (bottom panels). Dashed line at bottom panel combines absorption of donor and acceptor.
 Left panel: Othogonal dyad. Right panel: parallel dyad.}    \label{f:absorption}
  \end{center}
\end{figure}
We used absorption and fluorescence spectra of constituting rylene subunits (Fig. \ref{f:absorption})  to parametrize electron-vibrational couplings. The frequencies, displacement and anharmonicities  $\hat{H}_{vib}$ for stretching vibrational mode are obtained by comparing positions and magnitudes of progression peaks of absorption and fluorescence as explained in Ref. \onlinecite{galestianpour2017anharmonic}. The peak profiles and Stokes shift are used to fit the bath parameters  $\lambda$,  $\Lambda$.   Parameterizations are summarized in Table II.

The absorption line-shape of the dyad (Fig. 3 bottom) indicates the transport regime.
The weak coupling absorption of orthogonal case can easily be reconstructed by adding the donor and acceptor absorption lineshapes (dotted line in Fig 3 bottom left). The analysis of absorption spectra thus does not yield non-zero value of $J$ here. Larger couplings of parallel arrangement suggest the departure from F\"{o}rster transport theory and explain the differences between the absorption of the dyad  and combined  donor and acceptor lineshapes (shown as dotted line in Fig. \ref{f:absorption} bottom right). Dyads absorption can thus be used  for a rough estimate of the coupling within the range of 150 - 250 cm$^{-1}$ (consistent with the QC evaluations previous section \ref{s:Coupling}) where
 the vibronic dimer model can only be successfully fitted.

\subsection{Dynamics of transient absorption - orthogonal  dyad}

\begin{figure}[]
  \begin{center}
    \includegraphics[width=\linewidth]{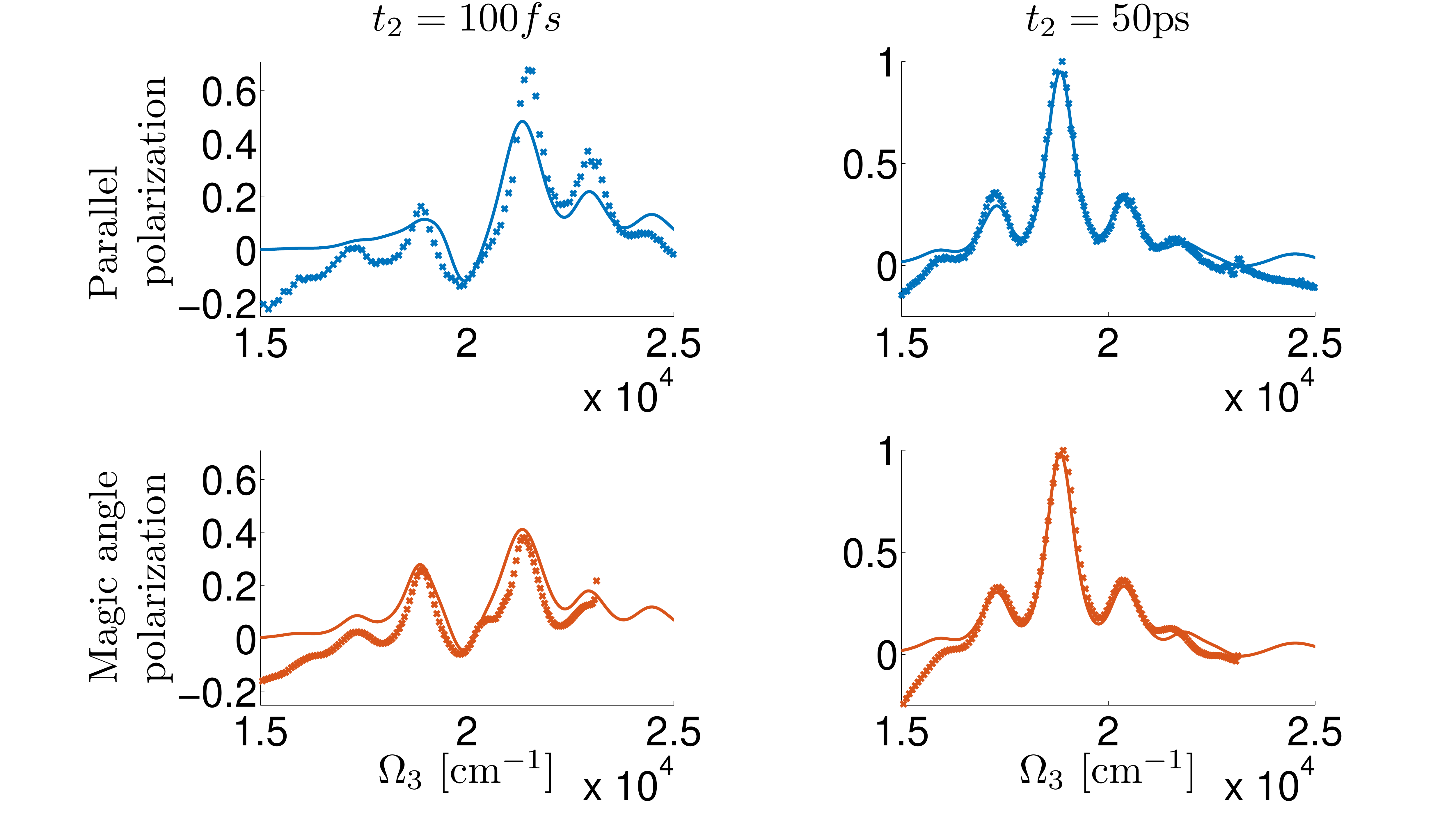}
  \caption{
  Transient absorption of orthogonal dyad in all parallel (top) and magic angle polarization (bottom) for early ($t_2 = 100$ fs left) and long waiting times ($t_2 = 50$ ps right).
  Experiment   (crosses) vs. simulation (solid line) using model parameters of Fig 4.  }
    \label{f:excitation}
  \end{center}
\end{figure}

\begin{figure}[]
  \begin{center}
    \includegraphics[width=\linewidth]{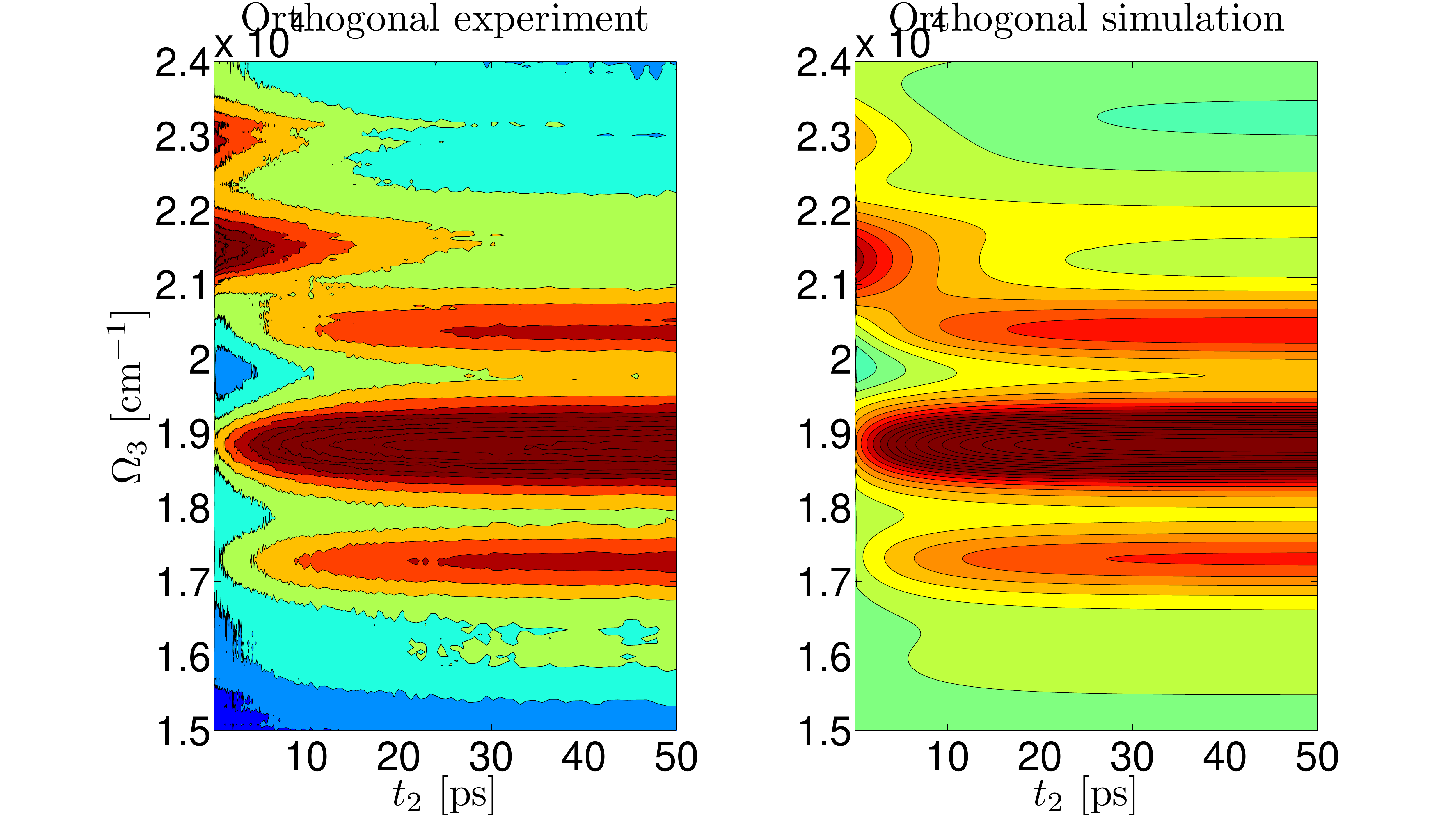}
     \includegraphics[width=\linewidth]{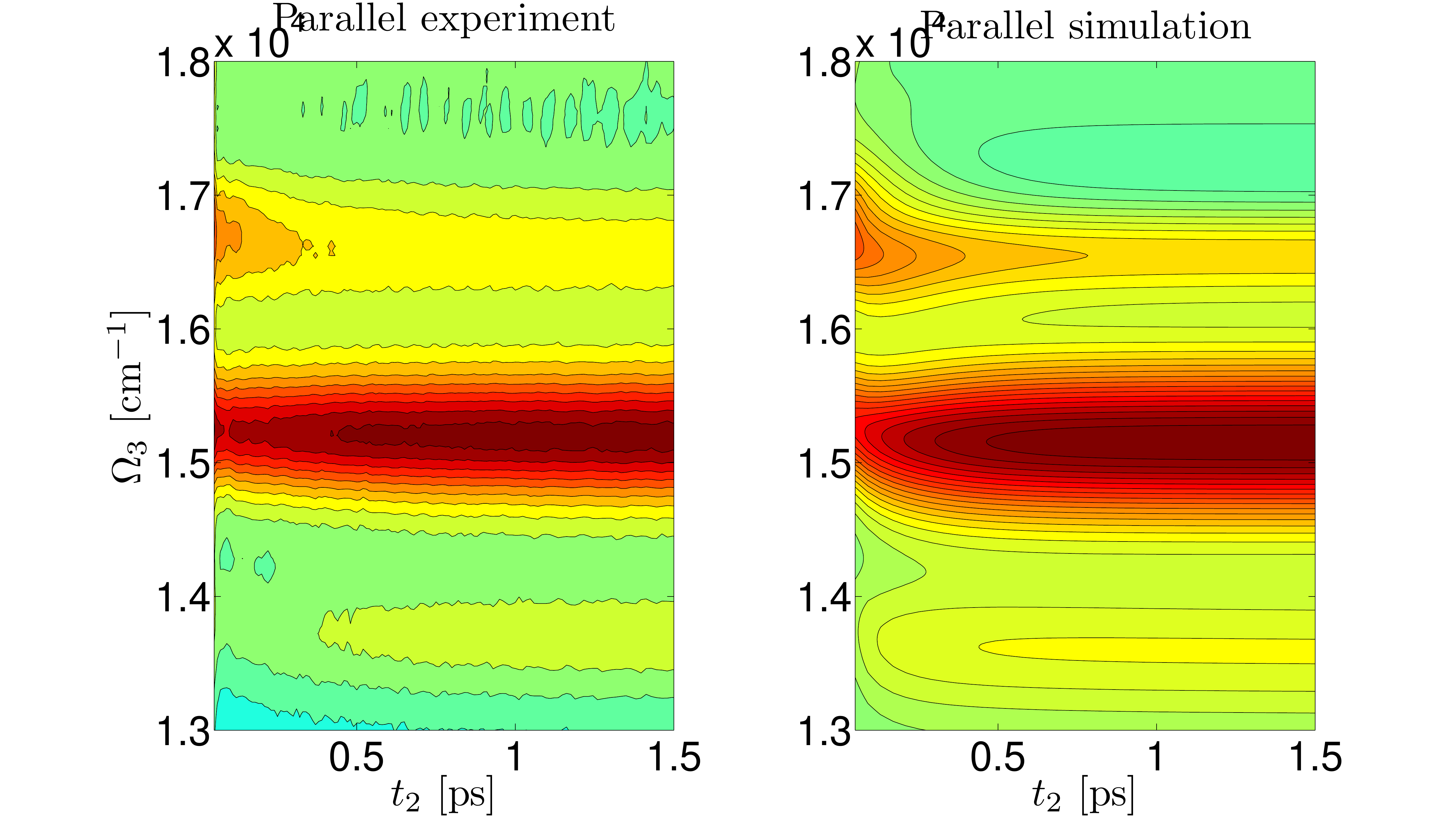}
  \caption{Delay time $t_2$ and detection frequency $\Omega_3$ dependencies of transient absorption signal of orthogonal (top panels) and parallel dyad (bottom) Left: experiment, Right: Simulations with $J=16$cm$^{-1}$ (orthogonal))  and $J=180$cm$^{-1}$ (parallel). Other parameters are summarized in Table II.  }
    \label{f:pump_probe}
  \end{center}
\end{figure}
\begin{figure}[]
  \begin{center}
    \includegraphics[width=\linewidth]{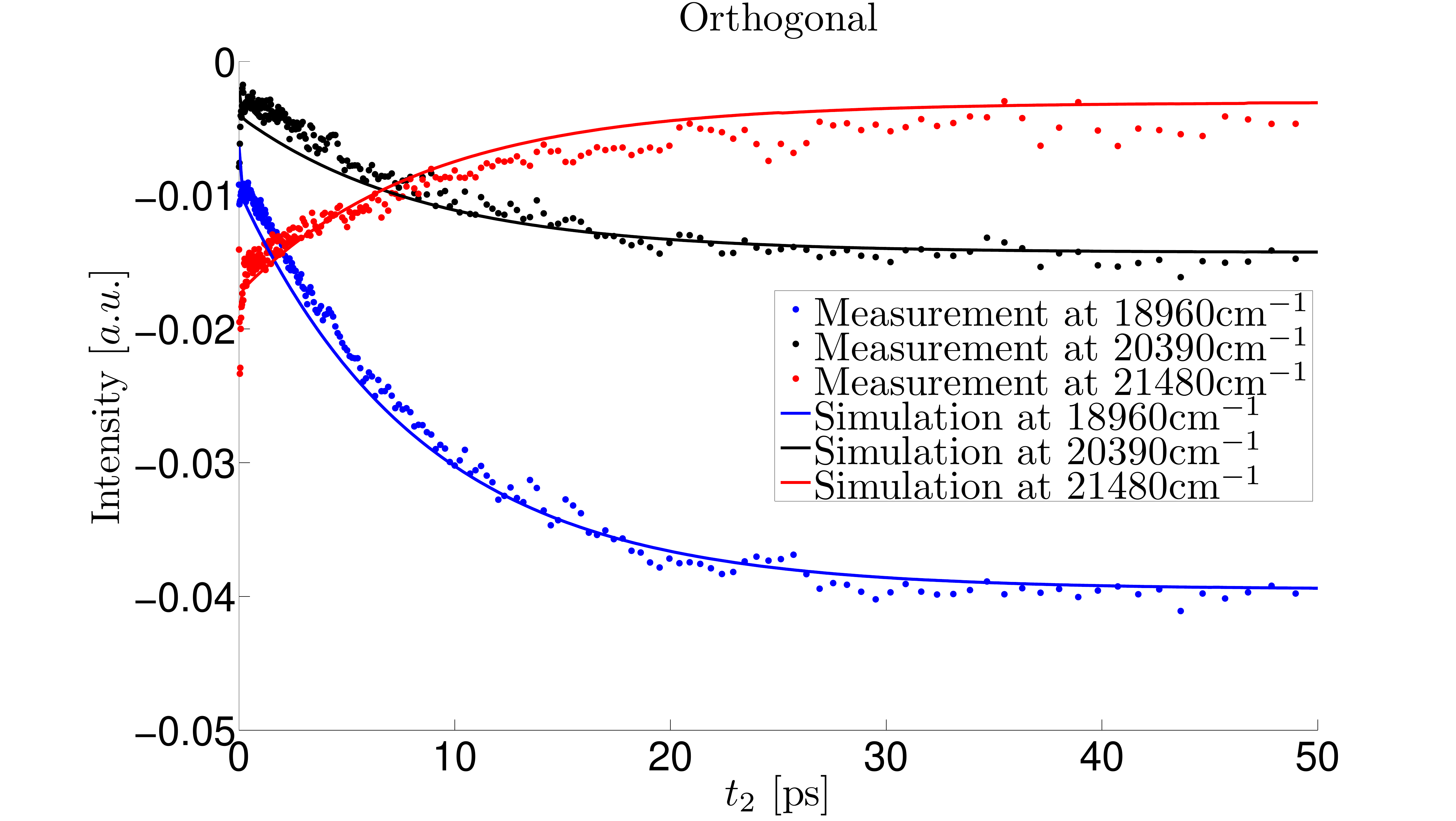}
  \caption{Trace of relaxation at slices of  indicated detection frequencies for orthogonal dyad, magic angle polarization set-up. Circles are experiment, solid line is simulation.}
    \label{f:cuts_orto}
  \end{center}
\end{figure}

We have  revisited the transient absorption of orthogonal dyad \cite{langhals2010forster,megerle2009sub} to
study time profiles of relaxation and its  polarization and frequency dependencies.
We compared  two pulse polarizations, all parallel pulses (Fig. \ref{f:excitation} top) and magic angle \cite{hamm2011concepts} geometry (Fig. \ref{f:excitation} bottom),
which represent different combinations, $\propto {\mathcal R}_{DD}+{\mathcal R}_{D\rightarrow A}$  and $\propto 3{\mathcal R}_{DD}+{\mathcal R}_{D\rightarrow A}$ , respectively, of donor
${\mathcal R}_{DD}$ and transport ${\mathcal R}_{D\rightarrow A}$ responses (assuming excitation at the donor).
 At early waiting times (Fig. \ref{f:excitation} left) the donor signal (${\mathcal R}_{DD}$) is prevailing and  the intensity of signal is different for the two polarizations.
 For the long waiting times we observe transport signal from acceptor ${\mathcal R}_{D\rightarrow A}$ with the same weight for the both polarization schemes and the amplitude difference is diminished.
 We conclude that the signals are consistent with theory.
 In the rest of this section, we opt for the magic angle  as we focus on transport, in addition to the general advantages of this set up (smaller sensitivity to orientational diffusion \cite{hamm2011concepts}).

Narrowband  excitation limits the observation window to delay times larger than 100 fs. We have not observed direct indication
of vibrational relaxation, nor significant excitation frequency dependencies in this window. We thus conclude the the stretch mode is mostly relaxed on this timescale
consistently with parametrization of Table II.
Below we report representative   spectrograms  (Fig. \ref{f:pump_probe}) for excitation in donor absorption region (465nm or $\Omega_1 = 21505$cm$^{-1}$).
The profile of relaxation at all three significant peaks at $\Omega_3=18960 $cm$^{-1}$ (representing e.g stimulated emission from vibrational ground of $|e_A\rangle$ state),
 $\Omega_3=20390 $cm$^{-1}$ ( representing resonances of vibrational ground of $|e_D\rangle $ with single vibrational excitations $|e_A\rangle$) , and $\Omega_3=21480 $cm$^{-1}$
 (resonances of single vibrational ground of $|e_D\rangle $ and double vibrational excitations of $|e_A\rangle$ ) is single exponential
 with similar lifetimes  $\tau_{18960}= 9.4 $ps and $\tau_{20390}= 9.3$  ps and $\tau_{21480}= 9.3$ps which represent vibronic resonance between donor's vibrational ground and acceptor with vibrational relaxation.
 This is the bottleneck of the dynamics, subsequent vibrational relaxation is too fast to modify the single-exponential character of transport.
Thus, the orthogonal dyad shows  all signatures of  F\"{o}rster type of transport limited by excitonic transport, with little indication of vibrational dynamics in the observation window.
The previously reported discrepancies can be explained by the difficulties with microscopic evaluations of coupling addressed in the previous section of the present publication.
The present vibronic simulations reproduce the transport rates for $J \approx 16$ cm$^{-1}$, and they reproduce the whole spectrum confirming the applicability
of our machinery to F\"{o}rster regime.

\begin{figure}[]
  \begin{center}
  \includegraphics[width=\linewidth]{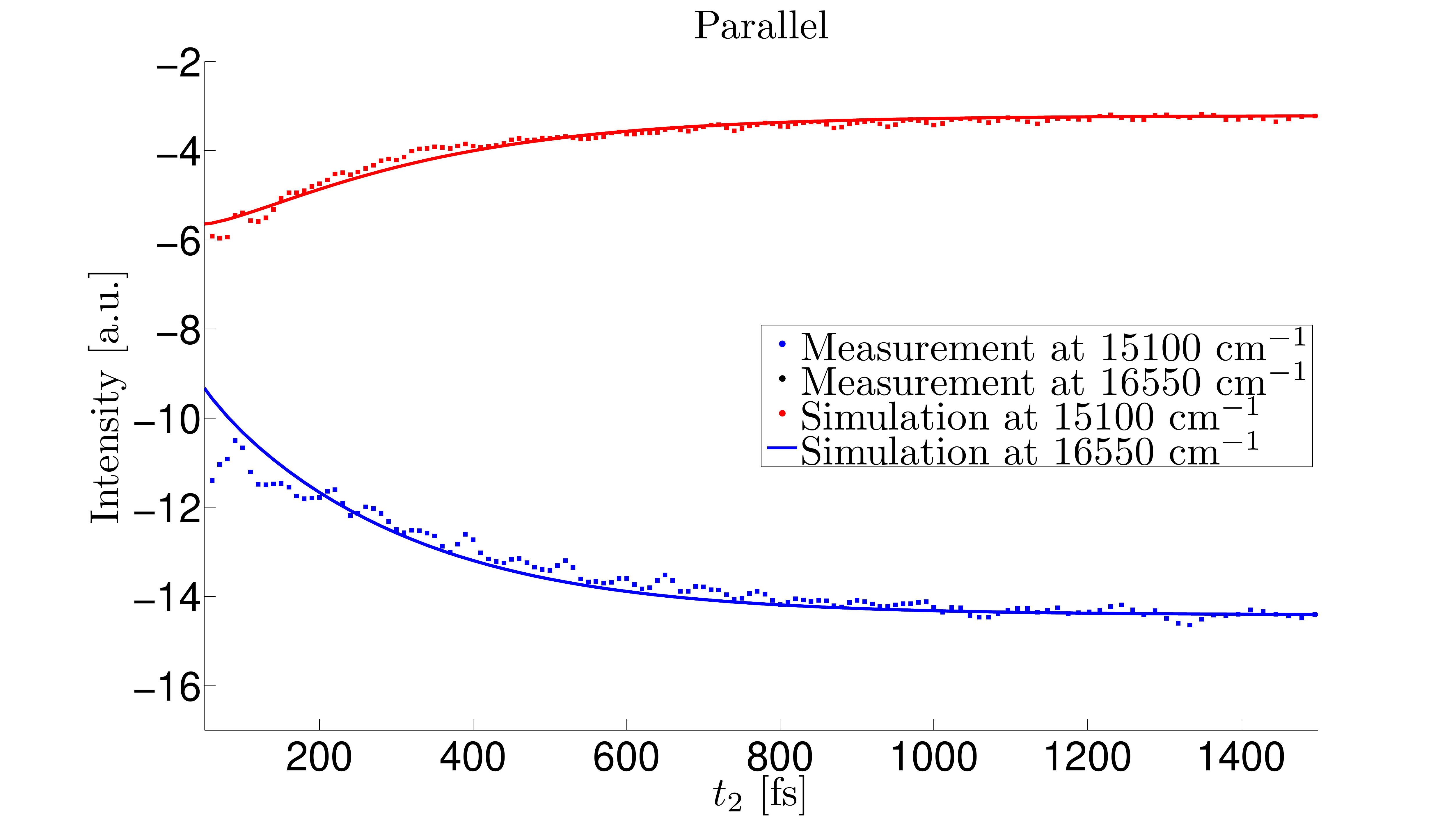}
  \caption
  {Comparison of simulated transient absorption of the parallel dyad to experimental data for resolved donor and acceptor peaks. Circles are experimental data, and solid line is numerical simulation.}
      \label{f:cuts_para}
  \end{center}
\end{figure}

\subsection{Dynamics of transient absorption - parallel dyad}

The transient absorption dynamics for the parallel dyad (Fig \ref{f:pump_probe} right) is polarization independent, detection frequency dependent, and much faster
as a result of much stronger coupling between the rylenes as compared to the F\"{o}rster regime of the orthogonal dyad treated  in the previous section.

For a more detailed analysis, we follow time evolution of two dominant spectral features: peak at 15100 cm$^{-1}$ representing acceptor principal transition and
peak centered at 16550 cm$^{-1}$ representing resonance of  donor principal transition with first vibrational progression of acceptor.
The delay time profile (Fig \ref{f:cuts_para}) of former peak is essentially exponential, the latter shows a more complicated dynamics.

\begin{figure}[]
  \begin{center}
 \includegraphics[width=\linewidth]{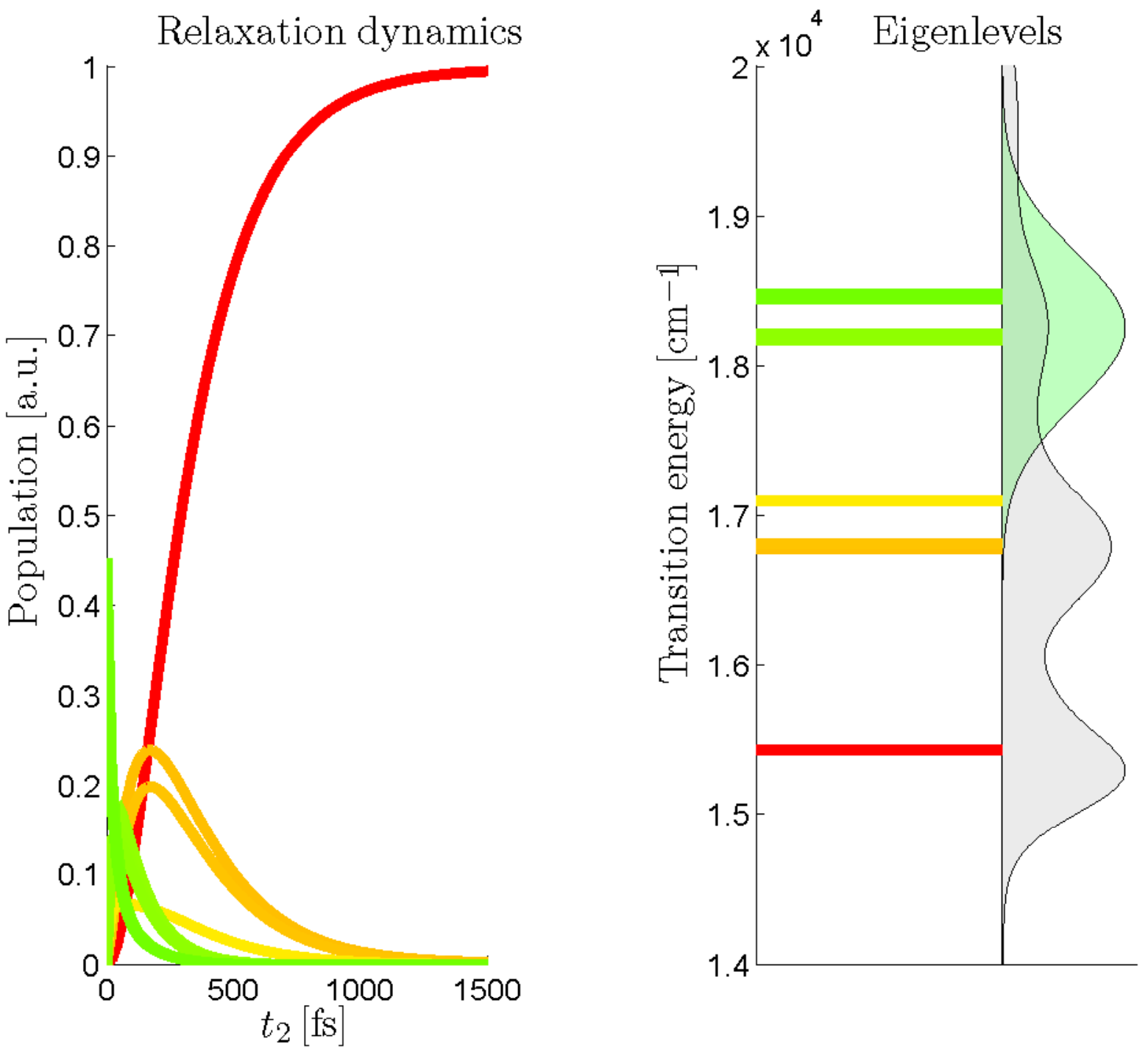}
  \caption{Relaxation in PD dyad. Left: Green function solution to population dynamics, initial condition
is pulse excitation at 548nm. Right: Eigenenergy ladder and relation
to absorption spectrum (gray) of the dyad. Pulse excitation marked green.  }
    \label{f:level_scheme}
  \end{center}
\end{figure}

To understand better the underlying transport dynamics we have unravelled the signal into the evolution of density matrix by using numerical simulations.
Simulations show beatings of signal at cca 20 fs timescale. This timescale is below the resolution of measurement and we thus averaged these beatings out.
The resulting pattern fits well with the experiments in Figs. \ref{f:pump_probe} and \ref{f:cuts_para} for parameters of Table II (obtained from fitting Fig. \ref{f:absorption}) and resonance coupling $J\approx 180$ cm$^{-1}$.
The electronic-vibrational relaxation pathways through the energy ladder are shown in Fig. \ref{f:level_scheme}. The dyad is primarily excited by a narrowband pulse in the green part of spectrum, i.e. into
the first progression of donor. It is quickly damped (no direct sign in spectrum)  into three "orange" levels, which optical properties are responsible for the complex dynamics of 16550 cm$^{-1}$ peak around 200 fs.
Finally, the relaxation is complete at the acceptor level and the densities and transient absorption lineshapes become stable after $t>1$ ps.

\section{Conclusion}
\label{s:Conclusion}

We have developed vibronic model of excitonic transport in rylene dimers
applicable to a variety of geometric arrangements and transport regimes of the dyad. In particular, we addressed
the absorption, fluorescence and  the transient absorption probes of donor-acceptor dyads in the orthogonal and parallel spatial arrangements.

The F\"{o}rster picture of transport dynamics was, in essence, confirmed for the orthogonal dyad,
by proving absorption spectra additivity, exponential delay-time profiles of transients and their independence on the detection frequency.
Previously reported inconsistencies between transport data and microscopic DFT parametrizations should thus be untangled from the latter end.
We overcame the arguably failing normal mode approach to thermal fluctuations by sampling the molecular geometries using MD with solvent
and statistically analyzed coupling distributions. We have found fairly complex fluctuation statistics which allows us to make a quantitatively adequate
 estimates of the effective intermolecular couplings.
A minor overestimate  may still be attributed to the neglect of solvent induced screening effects in  DFT calculations.
The spectra of parallel dyad have been measured and successfully simulated using the same methodology. The detection frequency dependencies witnessing the interplay of excitonic transport
and vibrational relaxation allow us to track the relaxation pathways within the dyad.

Our work is leaving several challenges for the future research.
The present  narrow band excitation  limits the direct view into sub 100 fs relaxation typical for dominant stretch.
The 2D electronic measurements  (scaling the pulse durations down towards 10 fs) will provide a more reliable  parametrization of the vibrational timescales.
Pulse polarization dependencies of the transients from orthogonal dyad manifested the transport between mutually orthogonal excitons.
Similar polarization effects in certain higher spectroscopies may be even used to probe directly the geometry dynamics \cite{sanda2011novel,mann2014probing}.
Elementary account for these dynamical dipoles (and, thus, also couplings) is possible by a stochastic (i.e. high temperature) extension of the present vibronic model in the spirit of Refs. \onlinecite{sanda2010exciton,sanda2011novel}. More sophisticated (finite temperature) hamiltonian vibronic dynamics with
dynamical dipoles and correct MD/DFT microscopic parametrization would be rather difficult challenge, but eventually widely significant for the tests of the transport theory.

\section*{Acknowledgements}
  J.H.  acknowledges
funding by the Austrian Science Fund (FWF): START project Y 631-N27. F.{\v{S}}., V.S., T.M. and V.P.
acknowledge the support by Czech Science Foundation (Grant No. 17-22160S). V.S. acknowledges the support by GAUK (Grant No. 1162216).
\bibliography{mybib}

\end{document}